\documentstyle[11pt,newpasp,twoside,epsf]{article}
\def\edcomment#1{\iffalse\marginpar{\raggedright\sl#1\/}\else\relax\fi}
\reversemarginpar

\begin{document}
\title{GBT Observations of Very Low Mass Binary Millisecond Pulsars:
A Search for Eclipses}
\author{David J. Nice}
\affil{Physics Department, Princeton University, Princeton, NJ 08544 USA}
\author{Ingrid H. Stairs}
\affil{Dept. of Phys. \& Astron., U.\,B.\,C., Vancouver BC V6T 1Z1 Canada}
\author{Zaven Arzoumanian}
\affil{USRA, LHEA, NASA-GSFC, Greenbelt, MD 20771 USA}


\begin{abstract}
We searched for eclipses of two
millisecond pulsars, PSR J1807$-$2459 and PSR B1908+00. These pulsars
are in very
low mass binary systems with
orbital parameters similar to those of eclipsing binaries.  Observations
were made with the GBT at frequencies as low as 575\,MHz.
No eclipses were detected in either system.  Observations of 
well-established eclipsing binary J2051$-$0827 found eclipses
to be substantially weaker than previously seen, with the
pulsar detected throughout the eclipse region at 575\,MHz and
with an electron column density an order of magnitude smaller
than previously measured.
\end{abstract}

\section{Introduction}

Very low mass eclipsing binary millisecond pulsars occupy a unique niche in
pulsar parameter space, between accreting X-ray binaries and
isolated neutron stars.  
In these systems, mass loss in the secondary is driven either
by Roche Lobe overflow or by winds induced by heating of the 
secondary by pulsar irradiation.  The resulting mass flow eclipses
the pulsar signal at inferior conjunction.

At this conference, the
results of two observational studies related to eclipsing binary
pulsars were given.  First,
multifrequency observations of orbital
and pulse-phase variability of eclipsing binary
pulsars B1744$-$24A and B1957+20 were presented.  This work will
be published elsewhere.  Second, a search
for eclipses in very low mass  pulsar binaries not
previously seen to eclipse was
described.  This paper summarizes the findings of the latter
experiment.

%
%

\section{Motivation}

There are more than a dozen known very low mass eclipsing binary pulsar
systems.  
There are also several pulsar binaries with similar orbital characteristics
but in which no eclipses have been detected
(see Table 1 of Paulo Freire's contribution to these proceedings).
The mass functions of the non-eclipsing systems are moderately
smaller than those of the eclipsing systems.
The question arises:  why do these systems not
exhibit eclipses?  There are three possibilities:
\begin{enumerate}
\item These systems are qualitatively different from eclipsing binaries.
Perhaps the pulsar irradiation is insufficient to drive mass loss
in the secondaries.  (The spin-down luminosities of this project's target
pulsars
are not known.)  Or perhaps the low mass functions indicate low secondary
star masses which, for whatever reason, are less prone to mass loss.
\item These systems are similar to eclipsing systems, with secondaries
nearly filling their Roche lobes, but  mass flow is
``turned off'' at the present epoch.  
In this scenario, eclipsing systems can become non-eclipsing systems,
and vice versa, perhaps on a time scale of years.  
Orbital elements of 
eclipsing systems exhibit small perturbations on timescales of years, 
but it is hard to see how they would influence the mass flow.  

\item These systems are not different than eclipsing systems, but our
viewing geometry is not favorable for detecting eclipses.  Their
orbital inclinations are much less than
90$^\circ$, i.e., they are not viewed ``edge on.''  This is consistent
with the low mass functions of these binaries if there is a relatively
narrow range of secondary masses among the low mass binaries.   
(See the discussion by 
Paulo Freire in these proceedings).

\end{enumerate}

It is a common property of very low mass eclipsing systems that
the eclipse characteristics depend on frequency.  For example, the
eclipse lengths, $\Delta T$,   of PSRs B1744$-$24A and B1957+20 scale
inversely with frequency, $\nu$, as $\Delta T \propto \nu^{\beta}$,
where $\beta=-0.63\pm0.18$ and $-0.41\pm 0.09$, respectively (Nice
et al. 1990; Fruchter et al. 1990).  Since eclipses observed at
low frequencies are extended
along the orbital plane, it is reasonable to
suppose that they are extended off the orbital plane as well, so
that, 
for lines of sight far from the plane (i.e., at low inclination)
eclipses might be detectable only at
low frequencies.

Observations of eclipsing binary PSR~J2051$-$0827 provide further
evidence that eclipses are more dramatic, and hence easier to detect,
at low radio frequencies.  Past observations of this pulsar found
no change in flux density, and only small dispersion delays, at 1400\,MHz
at inferior conjunction, while the system was completely
eclipsed at 660 MHz (Stappers et al. 1996; but see below).

\section{Observations}

We made low frequency observations of very low mass binary pulsars
PSR J1807$-$2459 and PSR B1908+00.
Their orbital periods 
are
1.7 and 3.4 hours, respectively, and companion masses are around
0.01 and 0.02\,M$_\odot$ under conventional assumptions.
Previous observations of these systems at 1400\,MHz
did not show eclipses (D'Amico et al. 2001, Deich et al. 1993,
Ransom et al. 2001).  
Observations of eclipsing binary pulsar 
PSR~J2051$-$0827 were made at some of the same epochs
as the other
pulsars.  While not originally a target of the eclipse-search project, 
these data are relevant to it (see below).

The pulsars were observed with the 100\,m
Robert C. Byrd Green Bank Telescope (GBT) at numerous frequencies
between 575\,MHz and 1660\,MHz.  The superior gain, frequency
agility, and tracking ability of this telescope make it
ideal for low-frequency observations of compact binary pulsars.
In a typical session, a
pulsar would be continuously observed for several hours---often
a complete orbit or longer.  Data were collected by
the Spectral Processor, a Fourier transform spectrometer.
Spectra were folded on-line modulo the
pulse period over intervals of 0.5 to 5.0 minutes. 
Off-line, the folded spectral data were de-dispersed, and 
pulse times of arrival were extracted using conventional techniques.  

\section{Results}

We found no evidence for eclipses of PSR J1807$-$2459 or PSR B1980+00.  
Eclipses would have been detected either by absence or systematic
reduction in flux around inferior conjunction or, in 
case of partial eclipses, by systematic delays
of the pulsed signal as it passed through the
ionized eclipsing medium.  Neither of
these observational signatures was detected.

The data are shown in Figures 1 and 2.  
Residual pulse arrival times are shown after
subtracting a model for the orbit.  The ascending
node is at orbital
phase $\phi=0$;
eclipses
would be at $\phi=0.25$.
Both pulsars are visible at $\phi=0.25$,
at frequencies
as low as 575~MHz (J1807$-$2459) and 820~MHz (B1908+00).

Limits on the electron column density of ionized
eclipsing material can be
derived from limits on the pulse time of arrival delays, $\Delta t$,
at $\phi=0.25$.  For J1807$-$2459, we estimate that
$\Delta t \la 20\,\mu$s, at 800\,MHz at conjunction, so the maximum
excess column density is $\Delta{\rm DM}=0.003\,{\rm pc}\,{\rm
cm}^{-3}=1.0\times 10^{16}\,{\rm cm}^{-2}$.  For B1908+00, $\Delta t
\la 200\,\mu$s at 820\,MHz yields $\Delta{\rm DM}=0.03\,{\rm pc}\,{\rm
cm}^{-3}=1.0\times 10^{17}\,{\rm cm}^{-2}$.  
These are both smaller than, for example, the past eclipses of
J2051$-$0827, for which
$\Delta {\rm DM}\approx 3\times 10^{17}\,{\rm cm}^{-2}$
(Stappers et al. 1996, 2001).

But PSR\,J2051$-$0827 presents a puzzle.  Its eclipses are known to be
variable, but it has always shown complete eclipses at
low frequency at some point near conjunction
(Stappers et al. 2001).  Observations of this pulsar with the GBT are
shown in Figures 3 and 4.  Surprisingly, it is visible
throughout the eclipse region, albeit attenuated as much as 80$-$90\%.
Further, the maximum time delay at conjunction is no more than $\Delta
t \sim 50\,\mu$s at 585\,MHz, which gives a column density of
$\Delta{\rm DM} \sim 1\times 10^{16}\,{\rm cm}^{-2}$, an
order of magnitude less than earlier measurements.  Clearly the
eclipses are much weaker than previously observed.

\section{Conclusion}

Neither of the target pulsars exhibited eclipses.  The question of
whether the lack of eclipses is intrinsic to the binaries or whether
it is a geometric effect remains open.  The surprisingly
weak eclipses in binary J2051$-$0827, with at least an
order of magnitude less ionized material around the secondary
than seen in previous observations, adds to the mystery.  While
there is no evidence that these systems completely ``turn off'' (as in
scenario no. 2 above), it is clear that the eclipse depths
can vary greatly.

\acknowledgements

We thank Scott Ransom for discussing previous observations of some
target pulsars.  This work was supported by NSF grant 0206205 and by
an NSERC UFA and Discovery Grant.  The GBT/NRAO is
a facility of the NSF,
operated by Associated Universities, Inc.

\begin{figure}[t]
\plotone{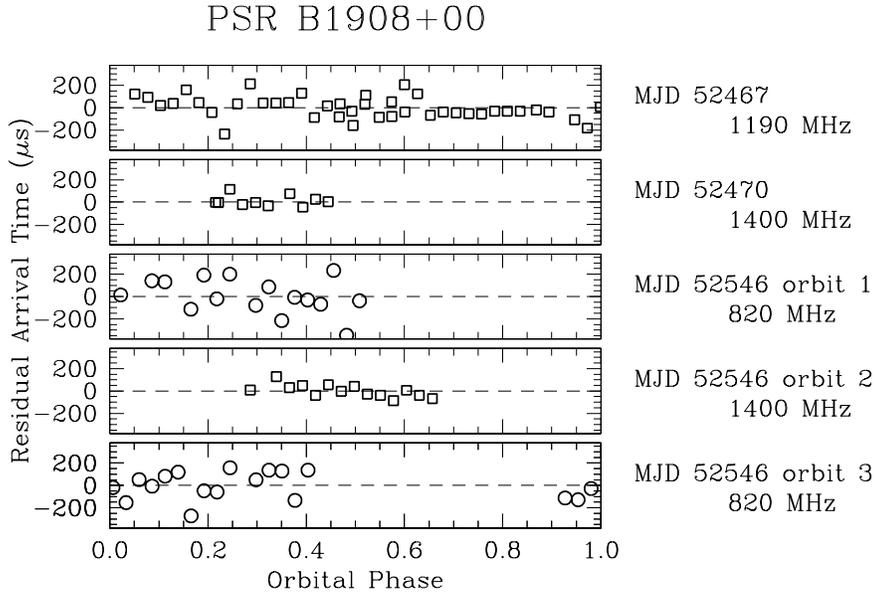}
\caption {\label{fig:1911}Residual pulse arrival times 
of PSR B1908+00, measured over five orbits
on three separate days.  Observing frequencies are 
indicated in the figure.
A partial eclipse would be indicated by arrival time delays,
or complete lack of signal, at orbital phase 0.25.
All GBT observations of this source are shown in
the figure.  Gaps and partial coverage of some orbits are the
result of telescope scheduling and/or instrumental
difficulties.}
\end{figure}

\vfill\clearpage

\begin{figure}
\plotone{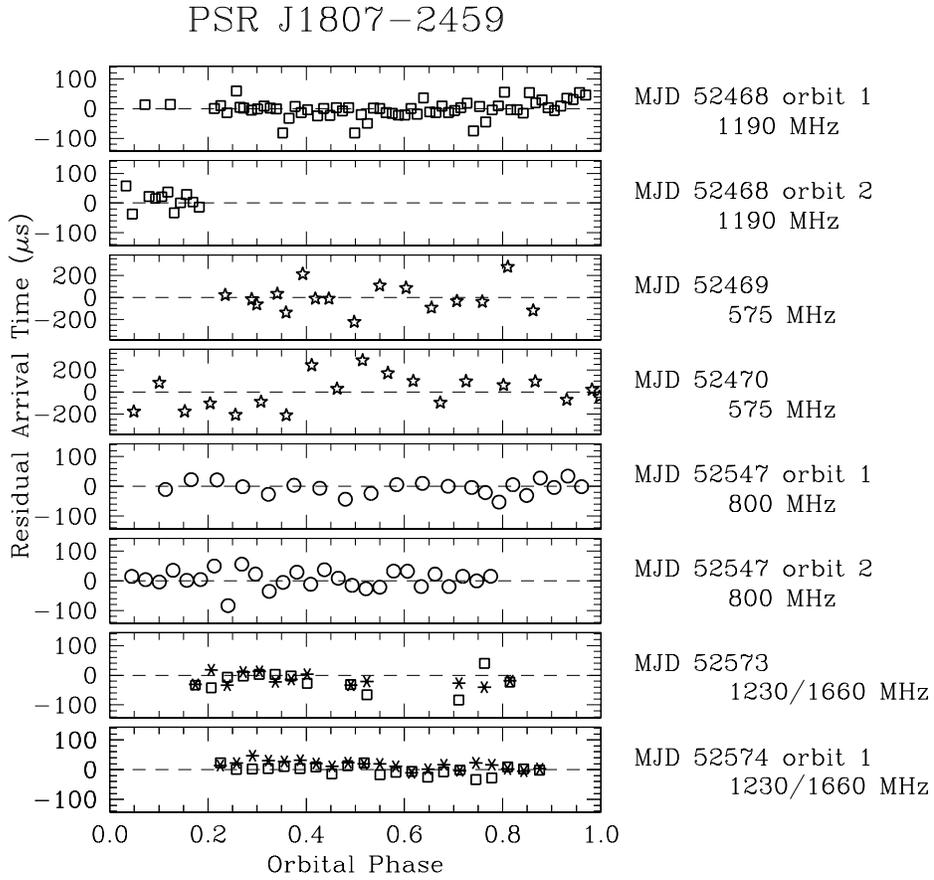}
\caption {\label{fig:1807}Residual pulse arrival times of
PSR~J1807$-$2459 measured over eight orbits on six separate
days.  Observing frequencies are indicated in the figure; 
simultaneous dual-frequency observations on MJDs 52573 and 52574
are indicated by squares (1230 MHz) and asterisks (1660 MHz).}

\end{figure}

\vfill\clearpage

\begin{figure}[t]
\plotone{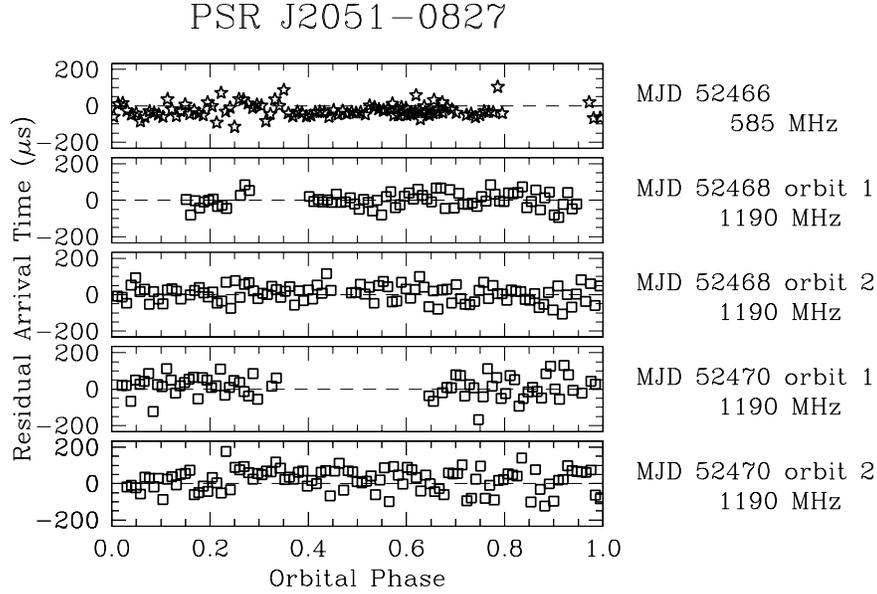}
\caption {\label{fig:2051}Residual pulse arrival times 
of PSR J2051$-$0827, measured over five orbits
on three separate days.  Observing frequencies are 
indicated in the figure.  Gaps in data indicate breaks
in data collection; they are {\it not} indicative of eclipses.}
\end{figure}

\begin{figure}[t]
\plotone{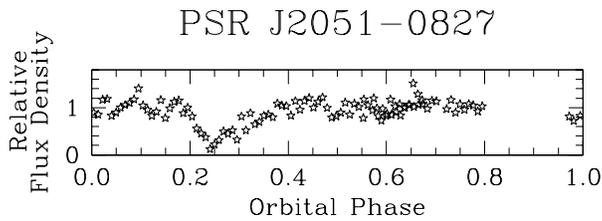}
\caption {\label{fig:2051b}Relative flux density
of PSR J2051$-$0827 over the course of the 585\,MHz
observations on MJD 52466.  In mid-eclipse, the 
flux density dipped as low as 10$-$20\% of its mean value, 
but it never went to zero (i.e., the pulsar
was always visible.)
}
\end{figure}



\end{document}